\documentstyle[multicol,aps,epsf]{revtex}

\begin{document}
\draft

\title{\bf Anisotropic Domain Growth of ANNNI Model at Low Temperatures} 
\author{Mookyung Cheon and Iksoo Chang}
\address{Department of Physics, Pusan National University, 
Pusan 609-735, Korea}
\address{and} 
\address{Department of Physics, Pennsylvania State University, 
University Park, PA 16802}
\maketitle

\begin{abstract}
We investigate the ordering kinetics for axial next nearest neighbor  
Ising (ANNNI) model in one and two dimensions by the multi-spin 
heat bath dynamical simulation. This dynamics enables us to overcome the 
pinning effect and to observe the dynamical scaling   
law for domain growth in the ANNNI model at zero temperature. 
The domain growth exponent is 1/2 isotropically both in the
ferromagnetic and the dry-(commensurate) antiphase. 
In the wet-(commensurate) antiphase, however,  
it is approximately 1/3 in the modulated direction, 
whereas it remains 1/2 in the non-modulated direction. 
We suggest that these exponent values are dictated by 
3 and 4 body diffusion-reaction processes of domain walls.
\end{abstract}


\begin{multicols}{2}

The phase ordering kinetics of systems quenched from a high temperature
disordered phase to a low temperature ordered phase has been widely
studied including the
domain growth law and the dynamical scaling behavior of correlation
functions \cite{Bray,RuBray,LMV,Iksoo}. 
It is generally accepted that the universal
behavior of the domain growth depends
on the presence of topological defects, the conservation law of 
the order parameter, and 
types of local energy
barriers\cite{RuBray,LMV,Iksoo,Shore,Rao}. 
Virtually all the systems studied thus far including anisotropic models such
as the axial next nearest neighbor Ising(ANNNI) model are believed
to have self-similar domains and an isotropic domain growth law
\cite{Kaski,Hassold,Ala,Redner}.
While previous studies of the ANNNI
model did observe somewhat anisotropic dynamic structure factors at low
temperature \cite{Kaski,Hassold},
the conventional belief is that the exponents characterizing 
the growth are isotropic. The previous work used single spin update algorithms
which cannot overcome the strong pinning effect at low temperature and leads
to sluggish dynamical behavior. Also, in accord with the prior expectation of isotropic
behavior, the total excess energy (in both direction) was used to find the
growth exponent \cite{Kaski,Hassold}.

The principal theme of this paper is a careful numerical study of the
dynamics of coarsening of the ANNNI model.
Our results indicate isotropic
growth at low temperature for a range of parameter values governing the
competition between the nearest and the next nearest neighbor exchange 
interaction in complete accord with previous expectations. 
However, we find in another part of parameter space striking evidence for  
anisotropic domain growth of self-affine domains 
with the exponent values (1/2 and 1/3) along
the different directions being
controlled by diffusion-reaction processes of the domain
walls (defects). These results are obtained using
a new multi-spin heat bath dynamics(MHBD) algorithm which allows one 
to overcome the strong pinning effects even at zero temperature. Our
work  provides the first vivid demonstration of anisotropic scaling
behavior in the coarsening problem and establishes the power of the
MHBD algorithm for effective
equilibration of this rich system.

The ANNNI model was first introduced to describe the equilibrium 
properties of spatially
modulated structures in magnetic and ferroelectric materials
where the commensurate-incommensurate (C-INC) transition exists
\cite{Elliott,Selke,Bak}.
The competing interactions in this model are ferromagnetic interactions
($J_1 > 0$) of nearest neighbor spins in all directions and 
antiferromagnetic interactions
($J_2 < 0$) of next nearest neighbor spins in the modulation direction 
and
give rise to a rich phase diagram: at low temperatures, depending
on the competition ratio $\kappa = -J_2/J_1$, the ferromagnetic phase exists
for $\kappa < 0.5$ and the (commensurate) modulated antiphase
($\cdot \cdot \cdot \uparrow \uparrow \downarrow \downarrow
\uparrow \uparrow \downarrow \downarrow \cdot \cdot \cdot$)
for $\kappa > 0.5$. This antiphase
consists of a wet ($0.5 < \kappa < 1$) and a dry phase ($\kappa > 1$), which
are separated by a line of wetting transitions as the temperature increases. 
On introducing the effect of thermal entropy, the phase diagram shows a variety 
of structures such as an incommensurate phase and devil's staircases.

Our MHBD algorithm considers
a square block of 4 spins
in 2D or a string of 4 spins in 1D, so that we update a spin-cluster
conformation among the 16 possible states.
Each state has its own probability in accordance with the Boltzmann weight
in the heat bath algorithm. One of the 16 states
is selected randomly according to their probability of occurrence and
consequently a new configuration of the 4 spins is obtained.
We first applied MHBD to the simple Ising model 
on a square lattice of linear size L=1000.
Simulations were carried out
at $T=0.1 J/k_B$ up to $10^4$ Monte Carlo step (MCS),
and averaged over 50 initial configurations. 
The expected scaling collapse of the correlation function is obtained and
is very consistent with that obtained with single spin
update algorithms. Moreover, the analysis from the excess energy,
the defect density, and
the correlation function provides a domain growth exponent of 1/2 . Thus MHBD
is confirmed to be an excellent approach for studying the ordering kinetics
at low temperatures.

We now consider a quenching of the linear ANNNI model, where the modulation
of ( $J_1 > 0$ and $J_2 < 0$ ) interactions exists precisely along the axis ,
from high $T$ to $T=0$
because an ordered state exists only at $T=0$. 
Right after a nucleation stage from the random initial states when $\kappa > 1$,
the system consists of 1($\cdot \cdot \cdot \downarrow \uparrow 
\downarrow \cdot \cdot \cdot$), 3($\cdot \cdot \cdot \downarrow 
\uparrow \uparrow \uparrow \downarrow \cdot \cdot \cdot$), 4( 
$\cdot \cdot \cdot \downarrow \uparrow \uparrow \uparrow \uparrow
\downarrow \cdot \cdot \cdot$), 5($\cdot \cdot \cdot \downarrow \uparrow 
\uparrow \uparrow \uparrow \uparrow \downarrow \cdot \cdot \cdot$), 
6($\cdot \cdot \cdot \downarrow \uparrow \uparrow \uparrow \uparrow
\uparrow \uparrow \downarrow \cdot \cdot \cdot$)-mers 
of up or down spins in the sea of dimers.
These 1, 3, 4, 5, 6-mers (called domain walls or defects) diffuse, annihilate 
and produce stable dimers by collisons as time elapses. 
Because the density of trimers would be higher
than any other $k$-mer except for the stable dimer, most of the excess energy
of this nonequilibrium state comes from trimers. When 
three trimers merge together ($\cdot \cdot \cdot \downarrow 
\uparrow \uparrow \uparrow \downarrow \downarrow \downarrow
\uparrow \uparrow \uparrow \downarrow \cdot \cdot \cdot$) 
, a monomer is produced with 4 stable dimers around ($\cdot \cdot \cdot
\downarrow \uparrow \uparrow \downarrow \downarrow  \uparrow
\downarrow \downarrow \uparrow \uparrow \downarrow \cdot \cdot \cdot$) 
and this monomer, when combined with a trimer, will result in two dimers 
eventually. 
This 3 body-collision and
annihilation of trimers is the dominant decay processes of domain walls and  
predicts the domain growth law
$L(t) \sim (t/\log t)^{1/2}$, where $L(t)$ is a characteristic 
length and the logarithmic correction originates from the fact that
D=1 is the critical dimension for the 3 body diffusion-reaction process
\cite{BLee,KKang}.
However, for $0.5 < \kappa < 1$, the domain walls are strongly pinned if their 
motion is controlled by a single spin update and thus such 3 body
annihilation or other decay processes of domain walls can never occur.
But, applying MHBD with a string of 2, 3 or 4 spins, 4 body annihilation
process of trimers can occur which then becomes 
the dominant decay process in this
regime: (3333) becomes (2433), (2442), or
(2622) with the same energy, and (2622) becomes 6 dimers with an
associated energy decrease $\Delta E = 4 J_1 + 8 J_2 < 0$ for
$0.5 < \kappa < 1$.  So, the rate equation for the density $A(t)$
of trimers, ${dA(t)\over dt} = - 4 A(t)^4 $, is obtained which  
immediately predicts the growth law 
$L(t) \sim  A(t)^{-1} \sim t^{1/3}$\cite{KKang}.

When 3 or 4 body collision of trimers
occur, 4-mer, 5-mer, and 6-mer can be considered as the resonance
of two trimers, three trimers, and four trimers. 
When 3 body collison of trimers is the dominant decay process among that of
$k$-mer domain walls, the density of $k$-mer domain walls will have a descending
order by 3, 1, 4, 5, 6-mers, whereas   
when 4 body collison of trimers is the dominant decay process, 
it will have a descending order by 3, 4, 5, 6, 1-mers.
Therefore, it is advantageous and important to look at the most and 
the next most dominant decay process of domain walls in the system,
which predicts 
the domain growth law of either $t^{1/2}$ or $t^{1/3}$ directly. 

We simulated a phase ordering of 1D-ANNNI model employing MHBD with blocks of 
either 2 or 4 spins for a system size 4000 up to $10^4$ MCS at $\kappa$=0.2,
0.4, 0.49, 0.51, 0.6, 0.9, 1.0, 1.01, 1.1, 1.5, 2.0, 3.0, 5.0, and 10.0.
The densities of $k$-mer domain walls, 
the excess energy as well as the correlation
function are calculated at each time and all
quantities are averaged over an ensemble of 2000 runs.
Figure 1 shows the densities of $k$-mer domain walls
at $\kappa=0.9$ and $\kappa=1.1$ in the (commensurate) antiphase. In the wet
phase ($\kappa=0.9$), the descending order in their densities 
are 3, 4, 5, 6, 1-mers 
and in the dry phase ($\kappa=1.1$), 
they are 3, 1, 4, 5, 6-mers. Therefore, the former
and the latter ought to have the domain growth law of $t^{1/3}$ and $t^{1/2}$, 
respectively. 
These results are self-consistently supported by our analysis of
the two-point correlation function.

\begin{figure}
\narrowtext
\centerline{\hbox{\epsfysize=2.0in \epsffile{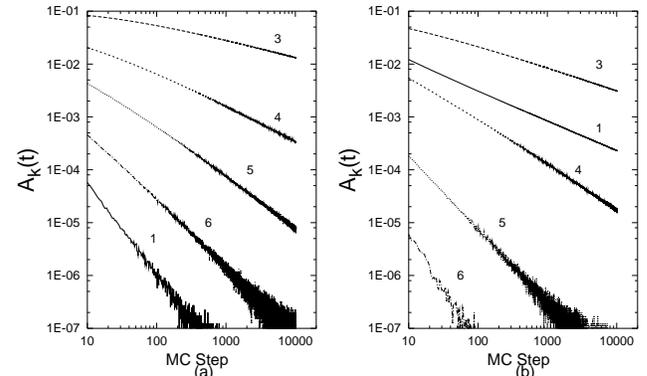}}}
\caption{\protect\footnotesize
The densities $A_k(t)$ of $k$-mer domain walls for the 1D-ANNNI model 
(a) in the wet phase ($\kappa=0.9$) : 3,4,5,6,1-mer from top and  
(b) in the dry phase ($\kappa=1.1$) : 3,1,4,5,6-mer from top. 
}
\end{figure}

\begin{figure}
\narrowtext
\centerline{\hbox{\epsfysize=2.0in \epsffile{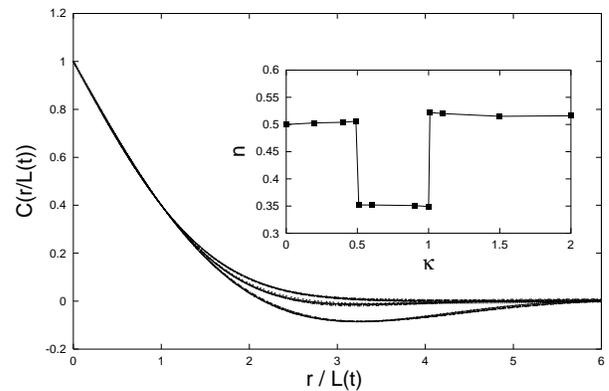}}}
\caption{\protect\footnotesize
The scaled correlation functions of 1D-ANNNI model in 
the ferromagnetic($\kappa = 0.4$), the dry($\kappa = 1.1$), and 
the wet ($\kappa = 0.6$) phase from top, which are collapsed for different
times $t =$ 80, 160, 320, 640, 1280, 2560, 5120, 10240 MCS at each
$\kappa$ values. Inset shows the growth exponents in each phases.   
}
\end{figure}

The two point correlation function $C(r,t)$ is defined 
as $C(r,t) = < S_i \cdot S_{i+r} \cdot (-1)^{r/2} >$ 
suitable for the antiphase , where $r$ is even number, and 
$C(r,t) = < S_i \cdot S_{i+r} > $ for the ferromagnetic phase. 
The growth exponent $n$ is evaluated by the decay of the excess
energy or the defect density and by the scaling collapse of 
the correlation function
$C(r/L(t))$, where $L(t)$ is obtained from $C(L(t),t)=1/2$ (see Fig. 2). 
For the ferromagnetic phase ($\kappa < 0.5) $, we obtain 
$n=0.5$ by fitting to a power law $L(t) \sim t^n$. 
For the dry-antiphase ($\kappa > 1$), $n \simeq 0.46 $ on fitting
to $L(t) \sim t^n $ while $ n \simeq 0.52$ on fitting to $L(t) \sim
(t/\log t)^n $. In order to detect the logarithmic correction, more
extensive simulations with system size $2\times10^4$ were performed
up to $6.5\times10^5$MCS at $\kappa=0.9$ and $1.1$ averaged over 300 samples.
The log-log plot of $L(t)$ versus $t$ at $\kappa=1.1$ is not a
straight line at longer times but
has an upward curvature. On the other hand, the log-log plot of
$L(t)$ versus $t/\log(t)$ shows a nice straight line.
This is fully consistent with the fact
that D=1 is the critical dimension of 3 body diffusion-reaction
processes \cite{BLee,KKang}. 
For the wet-antiphase ($0.5 < \kappa \leq 1$),
the growth exponent $n \simeq 0.35$ with a power law fitting, which is close to 1/3 
and consistent with the theoretical prediction \cite{KKang}.  
This is a rare realization of 4 body diffusion-reaction process whose dynamic 
scaling is first observed in our numerical simulations. 

For a 2D-ANNNI model, the ground state in the antiphase has 4 degenerate 
states (A,B,C,D). If the ground state at
two boundaries along the modulated ($y$) direction of the finite system
is fixed by A-state at one end and D-state at the other end, B and C-states 
can exist in between at equilibrium in the wet phase ($0.5 < \kappa < 1$). 
In the dry phase ($\kappa > 1$), the A-state meets the D-state directly
without the appearance of B and C-states. 
The domain patterns in the dry and the wet phases are quite different (see Fig. 3): 
For a dry phase, it 
looks like a typical domain growth pattern having 4 degenerate ground states.
But, for a wet
phase, it is elongated (growing faster) along the non-modulated ($x$) direction.
At $T=0$ the wetting transition, similar to that of liquid between air and substrate,
occurs at $\kappa = 1.0$, and there exists a line of wetting transition as the
temperature increases \cite{Rujan}. 

\begin{figure}
\narrowtext
\centerline{\hbox{\epsfysize=2.0in \epsffile{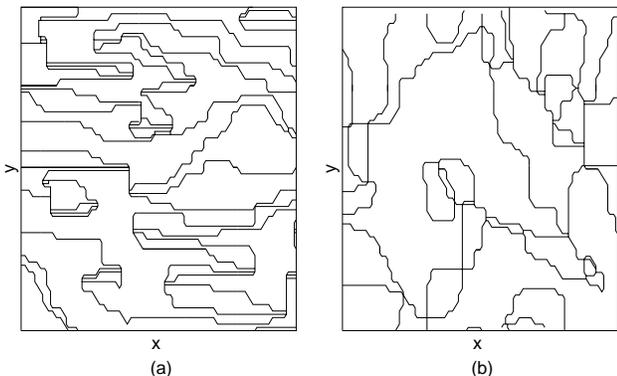}}}
\caption{\protect\footnotesize
The domain patterns of 2D-ANNNI model after 160MCS
(a) at $\kappa=0.6$ (the wet phase) 
(b) at $\kappa=2.0$ (the dry phase). 
}
\end{figure}

We performed MHBD simulations for ordering kinetics of the 2D-ANNNI model 
at $T=0$ for several $\kappa$ values ($\kappa$=0.2, 0.4, 0.6,
0.9, 1.0, 1.1, and 2.0) taking 4 spins as a square block. 
The system size is $2048 \times 2048$ and 30 samples are
accumulated up to $10^4$MCS.
MHBD could remove 
the pinning effect for $\kappa < 1$ so that
domains grow well in both the ferromagnetic and the antiphase. 
The two point correlation functions for the antiphase were defined
separately by $C_x(x,t)=<S_i \cdot S_{i+x}>$ in the non-modulated ($x$) direction and
$C_y(y,t)=<S_i \cdot S_{i+y} \cdot (-1)^{y/2}>$ in the modulated ($y$)
direction. The excess energy and the length of domain walls (defects) are
calculated separately for each direction. The densities of $k$-mers
along the $y$-direction are also calculated as in the 1D case. 
The growth exponents
are evaluated by $L_x(t) \sim t^{n_x}$ and $L_y(t) \sim t^{n_y}$
obtained from the scaling collapse of correlation functions and 
by the domain wall densities 
$\rho_x(t) \sim t^{-n_x}$ and $\rho_y(t) \sim t^{-n_y}$.

The growth exponents $n_x$ and $n_y$ are 1/2 at $\kappa=0.2$ and $0.4$
(ferromagnetic phase) and also at $\kappa=1.1$ and $2.0$ (dry-antiphase).
But, at $\kappa = 0.6, 0.9, 1.0$ (wet-antiphase), 
$n_x$ and $n_y$ are anisotropic:
$n_x \sim 0.55$ and $n_y \sim 0.39$.
One may wonder whether these values are transient and  
may both approach 1/2 (or 1/3
and 1/2 respectively) in the very long time limit. More simulations with
the larger system size $4096\times4096$ were performed up to $2\times10^4$MCS at
$\kappa=0.8$ and $1.0$. Further more, MHBD with $3\times3$ and $4\times4$
block cells were used under the same conditions. However, the exponents
$n_x \simeq 0.55$ and $n_y \simeq 0.39$ do not change significantly \cite{Matsu-1}. 
Hence one may also expect that the correction term must be
included in $L_x(t)$ and $L_y(t)$. We considered two kinds of corrections:
The first is $\rho(t)^{-1} \sim L(t) \sim t^n (1+t^{-\Delta})$ and
the second is $\sim t^n (\log t)^{\alpha}$. The
non-linear fitting of the former does not give consistent
results for different time intervals and $\kappa$ values.
However, the same fitting of the latter gives 
$n_x \simeq 1/2 , \alpha \simeq 1/2$ and $n_y \simeq 1/3 , \alpha \simeq
1/3$ for many different time intervals and $\kappa$ values.
So, the anisotropic domain growth laws in the wet-antiphase
are consistent with $L_x(t) \sim (t \log t )^{1/2}$ and
$L_y(t) \sim (t \log t )^{1/3}$.
Figure 4 shows the nice scaling collapse of correlation 
functions both for the non-modulated ($x$) direction and 
for the modulated ($y$) direction in the wet and dry phase. 
A similar collapse is observed for a plot of the domain wall density
(not shown).

\begin{figure}
\narrowtext
\centerline{\hbox{\epsfysize=2.0in \epsffile{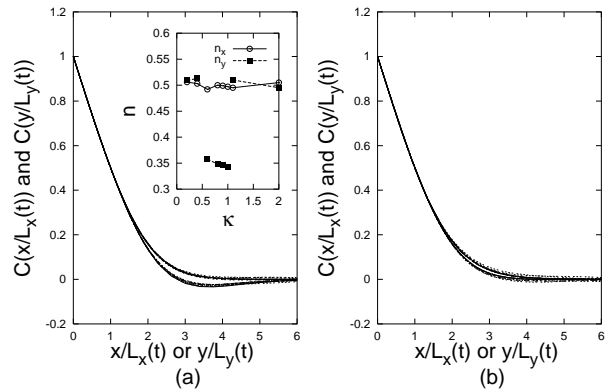}}}
\caption{\protect\footnotesize
The scaled correlation function of the 2D-ANNNI model for
non-modulated ($x$) direction ( top curve : $C(x/L_x(t))$ vs
$x/L_x(t)$ ) and for modulated ($y$) direction ( bottom curve :
$C(y/L_y(t))$ vs $y/L_y(t)$ )
(a) in the wet phase ($\kappa$=1.0) and (b) in the dry phase
($\kappa$=2.0) collapsed for different times $t =$ 40, 80, 160,
320, 640, 1280, 2560, 5120, 10240 MCS at each $\kappa$ values.
The similar collapse of correlation functions in the ferromagnetic
phase was also achieved.
Inset shows the growth exponents in different phases.
}
\end{figure}

The result from the densities of $k$-mers along the 
$y$-direction further supports and 
is consistent with the above
remarkable result because their descending sequence is 
3,4,5,6,1 at $\kappa=0.9$ and 3,1,4,5,6 at $\kappa=1.1$
in agreement with the 1D case (see Fig. 5). 
So, one may identify that the 4 (3) body diffusion-reaction
process of domain walls is important, as in 1D, as the dominant decay process  
in the wet (dry) phase of 2D-ANNNI model, which results in the dominant
domain growth law of $t^{1/3}~ (t^{1/2})$.  

\begin{figure}
\narrowtext
\centerline{\hbox{\epsfysize=2.0in \epsffile{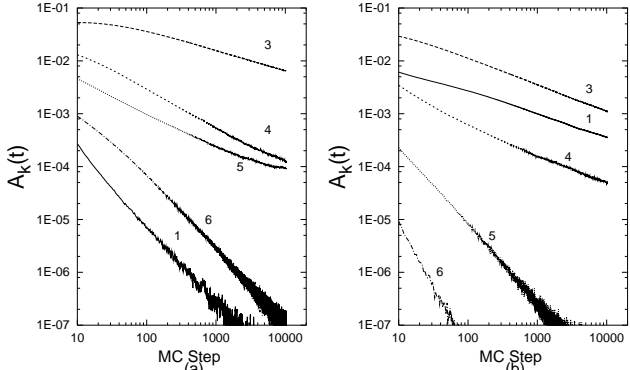}}}
\caption{\protect\footnotesize
The densities of $k$-mer for the 2D-ANNNI model.
(a) $\kappa=0.9$ (the wet phase) : 3,4,5,6,1-mer from top. 
(b) $\kappa=1.1$ (the dry phase) : 3,1,4,5,6-mer from top. 
}
\end{figure}

One may actually consider that the domain
growth at $T=0$ is possible since MHBD is adopted. 
However, the simulation by a single spin update, Glauber dynamics,
at $T = 0.4J_1/k_B$ in the wet phase 
gives the similar anisotropic growth exponents $n_x \simeq 0.56$ and
$n_y \simeq 0.40$. The two point correlation function along the
modulated $(y)$ direction has also the same dip as that from using MHBD.
Therefore, a single spin update dynamics can give the 
similar but less precise result for $T > 0$, 
since it severly suffers from 
the strong pinning effect at the low temperatures.

We have presented a summary of our results of detailed investigation of
the ordering kinetics of the 1D and the 2D-ANNNI model at $T=0$   
using the multi-spin heat bath dynamical simulation to overcome 
the pinning effect. Our simulations show that
the dominant domain growth exponent becomes 1/2 isotropically 
both in the ferromagnetic and dry-(commensurate) antiphase.
In the wet-(commensurate) antiphase, however,  
it is approximately 1/3 for the modulated direction, 
whereas it remains 1/2 for the non-modulated direction. 
The exponent values are explained by
3 and 4 body diffusion-reaction processes of domain walls (defects). 
Contrary to conventional belief regarding universality,
our data suggest an extremely unusual situation in which the
very nature of the domain growth law critically depends on the
ratio between the strength of competing interactions.

In the context of domain growth
dynamics of experimentally relevant systems, 
the chemisorbed adsorbate systems 
such as $\rm {O/Pd(110)}$ and $\rm {H/Fe(110)}$ are of
particular interest \cite{Rujan,Etrl,Sselke}.
Both systems exhibit a wetting phenomena in the $(3 \times 1)$ phase. 
It will be worthwhile to probe a temporal
evolution of structual factor for these systems 
experimentally, and compare with our results.

We are indebted to Jayanth Banavar, Yonathan Shapir and Daniel Hong  
for valuable discussions. 
We acknowlege the allocation of extensive CPU time by
the High Performance Supercomputing Center of PNU, and the grant  
(KRF-2000-DS0011) from Korea Research Foundation.

\end{multicols}

\begin{references}
\bibitem{Matsu-1}
We also used the cluster heat bath dynamics(CHBD) \cite{Matsu} with $2\times2$,
$3\times3$, and $4\times4$ block cell, which determines the spin values
of only one column of a block cell simultaneously by a transfer matrix method.
This CHBD gives virtually the same results as MHBD.
\end{references}
\end{document}